\documentclass[aps,prl,twocolumn,groupedaddress,showpacs]{revtex4}
\usepackage{graphicx}

\begin{document}


\title{Picosecond time-resolved two-dimensional ballistic electron transport}


\author{E.A. Shaner}
\email[]{eashaner@princeton.edu}
\author{S.A. Lyon}
\email[]{lyon@princeton.edu}
\affiliation{Princeton University Electrical Engineering
Department}


\date{\today}

\begin{abstract}
Time-resolved transport of ballistic electrons in a two-dimensional electron gas has been measured with a resolution of less than $5ps$.  This was accomplished by using picosecond electrical pulses to launch electrons from the emitter of a transverse magnetic focusing structure and optoelectronically sampling the collector voltage. Both plasma resonances and the ballistic transport signal are clearly resolved.  The transit time appears to be somewhat longer than expected from simple Fermi velocity considerations.

\end{abstract}

\pacs{71.45.Gm, 78.47.+p, 78.67.-n}

\maketitle


Time-dependent transport measurements in two-dimensional electron gas (2DEG) systems facilitate investigations of the charge distribution and its collective dynamics\cite{buttiker}.  Techniques implementing fast sampling oscilloscopes have been used to detect edge-magnetoplasmon excitations in the time domain at temperatures as low as $0.3K$ and magnetic fields as high as $13T$\cite{Ashoori,VK2}.  Typical time resolution achieved in these experiments was on the order of $100ps$. To measure the dynamic response of a mesoscopic device, a topic of significant theoretical interest\cite{buttiker,Igor}, one encounters additional experimental difficulties. Ballistic electron transport times in mesoscopic devices fabricated from AlGaAs/GaAs 2DEGs can easily range from a few to hundreds of picoseconds.  To fully characterize the device response in this regime, it is advantageous to provide picosecond scale \textit{excitation} and \textit{detection} while still working in the environment of low temperatures and high magnetic fields. In this letter, we report the direct measurement of picosecond time-resolved ballistic electron transport in a 2DEG transverse magnetic focusing device.  Our approach utilizes an ultrafast waveguide and integrated detector coupled to the device under test.

\begin{figure}
\centering
\resizebox*{7.5cm}{!}{\includegraphics{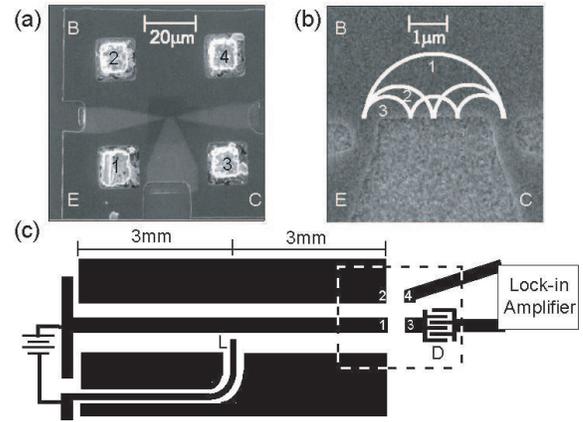}}
\caption{(a) 2DEG chip (processed concurrently with that used in the measurements) showing four $15\mu m\times 15\mu m$ GeNiAu contact pads with $10\mu m\times 10\mu m$ Indium bumps on top.  Final mesa is $\sim 85\mu m\times 85\mu m$ (b) Zoomed view of the point contacts and transport region.  Schematic focusing orbits for n = 1,2,3 are shown. (c) Waveguide sample layout.  Launch (pump) switch is labelled 'L', detector (probe) switch is labelled 'D'. Spacing between $15\mu m$ wide center conductor and ground planes is $30\mu m$. Dashed square shows where 2DEG chip is attached in final sample.} 
\label{fig1}
\end{figure}

    For high frequency investigations of 2DEG transport, it is convenient to have a way to distinguish the electron signal from other coupling between the device nodes.  Previous experiments investigating edge-magnetoplasmons have used gates to selectively deplete the 2DEG and thus determine its contribution to the overall signal\cite{VK2}.  In our experiment we chose to work in a transverse magnetic focusing geometry as shown in Fig.1a.  In a conventional experiment (DC), a small current between contacts $(1)$ and $(2)$, is passed through the emitter point contact (labelled \textit{E} in Fig.1b) and the voltage between contacts $(3)$ and $(4)$ is sensed across the collector point contact (labelled \textit{C} in Fig.1b).  By applying a perpendicular magnetic field, ballistic electrons injected into the base of the device can be deflected, by the Lorentz force, into the collector either directly or via skipping trajectories along the device boundary.  As the magnetic field is swept, this leads to a periodic change in collector voltage with peaks at fields meeting the condition $B=nB_{focus}$ for integer n where $B_{focus}=2\hbar k_{F}/eL$\cite{VanHouten}. Here $L$ is the point contact spacing and $k_{F}$ is the Fermi wave vector determined by the 2DEG density.  This signal is only observed in one direction of magnetic field as the opposite polarity deflects electrons away from the collector.  The periodicity in B along with the one sided spectrum provides a means to distinguish the ballistic electron signal from the background. 

    The focusing device of Fig.1a was fabricated from an $Al_{0.3}Ga_{0.7}As/GaAs$ modulation doped heterostructure.  The 2DEG formed at the interface was located $700\AA$ below the sample surface.  The device was patterned using electron beam lithography and a shallow etch to define the point contacts ($0.7\mu m$ wide with $3.7\mu m$ center-to-center spacing).  Alloyed Ni-Au-Ge contacts were made followed by a $1000\AA$ deep mesa etch to remove the 2DEG outside the region of interest.  This device isolation was essential as bulk 2DEG magnetoplasmons are known to couple strongly to waveguide structures\cite{Shaner}.  The processed device was without detectable carriers before illumination and had an electron density of $\sim 2.3\times 10^{11}cm^{-2}$ after illumination.  The estimated mean free path of the electrons in this device was $9\mu m$.  The illumination in this case originated from stray laser light (used for driving Auston switches) striking the device mesa.  When the laser was turned off, the carrier density remained stable and there were no noticeable changes when it was left on for the duration of the experiment.

A separate waveguide device was fabricated that utilized low temperature grown GaAs (LT-GaAs) Auston switches \cite{Auston}  for both pulse generation and detection.  Due to the short ($\sim 1ps$) recombination time in LT-GaAs, it is an excellent photoconductor for ultrafast applications\cite{LT}. The waveguide conductors were made using $200\AA$  Ti $/2000\AA$ Au in a liftoff process.  The launch (L) and detection (D) points on the waveguide are shown in Fig.1c, along with the relevant device dimensions. The waveguide substrate was a multilayer structure comprised of a $2\mu m$ LT-GaAs epilayer transferred onto a $200\mu m$ thick sapphire plate that was subsequently bonded to silicon.  Holes were etched completely through the silicon to hold single mode optical fibers that illuminate the switches\cite{Shaner}.  Indium bumps of $2\mu m$ height were evaporated on the connection points (enumerated in Fig.1 on both devices). The pieces were then brought together using a flip-chip bonder to make the electrical connections between the waveguide and the focusing device.  Epoxy was used to fill the remaining gap of $2-3\mu m$ between the two devices.

    Pump and probe optical fibers were attached to the back of the waveguide in order to illuminate the launch and detection switches. The sample was mounted inside a small superconducting magnet and measurements were performed at $4.2K$ in liquid helium. When a DC bias was applied across the launch switch, as shown in Fig.1c, an optical pulse sent down the pump fiber generated a lifetime limited photocurrent in the LT-GaAs.  This current produced a voltage transient on the waveguide. Using a fiber dispersion compensator and $250fs$ wide $790nm$ wavelength optical pulses produced by a mode locked Ti:Sapphire laser,  the minimum pulse width attained on a similar calibration  waveguide (same conductor spacing and substrate) was $1.8ps$.  On the detection side, an optical pulse received from the probe fiber electrically closed the detector switch and sampled the collector voltage with similar lifetime limited resolution. 

  To perform measurements, a DC bias of $-15V$ was applied across the launch switch while one end of the center waveguide conductor (connected to contact (1) at the other end) and contact (2) were held at signal ground.  The pump beam was chopped at a frequency of $11Hz$ allowing for differential lock-in detection of the voltage between the output of the detector switch and the opposing transport connection (4).  An optical delay line was inserted into the pump beam path and the delay between probe and pump was stepped by $1ps$ between magnetic field sweeps from $-330mT$ to $330mT$.  Time averaged optical powers of $300\mu W$ and $1mW$ were used for the pump and probe respectively.

Zero delay (t=0) denotes the time when the electrical pulse generated on the waveguide reaches contact (1) of the focusing device.  Its location (delay line position) was determined by careful measurements of the optical fiber lengths.  At the conclusion of this series of experiments, the fibers were removed from the device and the location of zero delay confirmed in the calibration waveguide.  Figure 2 shows magnetic field sweeps at selected delay times.  For times up to $t=30ps$, no focusing signal can be seen.  At $t=50ps$, the focusing signal is clearly developed and the focusing oscillations are seen to decay in amplitude for longer delay times.

\begin{figure}
\centering
\resizebox*{8.0cm}{!}{\includegraphics{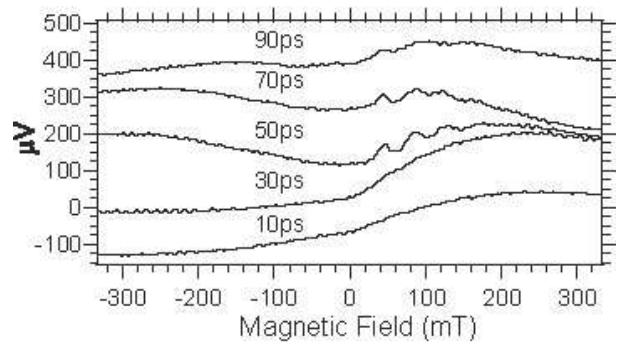}}
\caption{Magnetic field sweeps at various time delays between emitter excitation and collector probing.  Scans were offset for clarity.  For positive magnetic fields, the magnetic focusing oscillations appear strongly in the $t=50ps$ trace with approximately a $40mT$ period.} 
\label{newfig2}
\end{figure}

The time dependence of the collector voltage at zero magnetic field is shown in Fig.3a. An initial pulse is seen in the collector response that coincides with the expected arrival time of the electrical pulse at the emitter to within $1ps$.  We interpret this signal as coupling between the emitter and collector nodes.  Although this coupling interferes with extracting a transport signal, it is useful in that it gives both an indication of the time dependence of the emitter voltage and a reference point in the data for when the electrical pulse reaches the focusing device.  Even though pulses with $1.8ps$ width are launched on the waveguide, this signal is broadened to $5ps$ which serves as an upper limit on the exciting pulse duration.  Also, because the waveguide did not have an impedance matched termination at the left handed end (Fig.1c), reflections are seen at later times which could launch additional electrons.  However, there is a window of about $50ps$ in which there are no major reflections.  The structure in this zero field response is reproducible at all times (within $\sim 5\mu V$ noise level - about the thickness of the trace in Fig.3a) and the signal at times earlier than those displayed was zero.

\begin{figure}
\centering
\resizebox*{8.0cm}{!}{\includegraphics{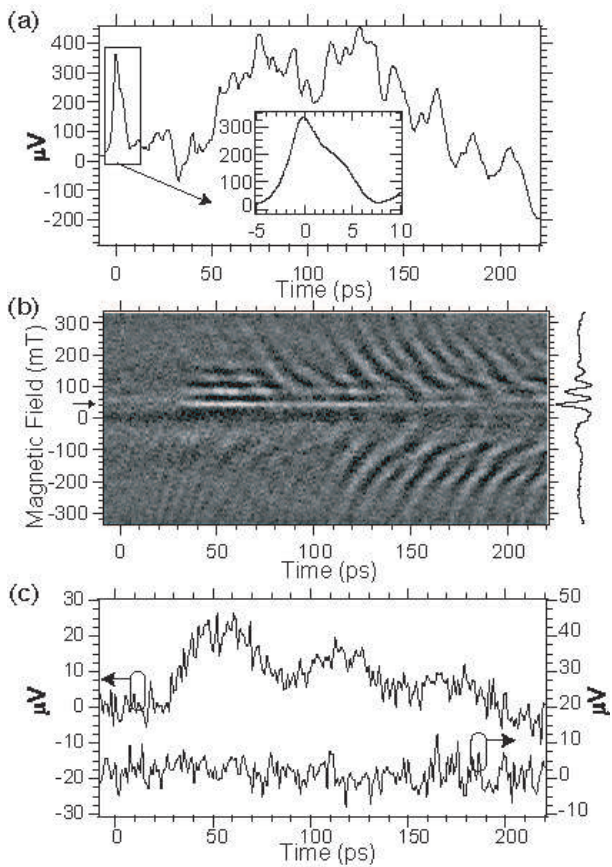}}
\caption{(a) Unfiltered collector response in the time domain at zero magnetic field.  The inset shows a high resolution time scan of the initial pulse about $t=0$.  (b) Filtered time resolved collector response presented in a greyscale image. Lighter areas correspond to electrons entering the collector. The arrow marks the location of the first focusing peak at B=0.043T.  A time average of the image data from time $t=25ps$ to $t=75ps$ is shown to the right of the image. (c) Upper trace shows a $5mT$ wide horizontal slice at the first focusing maximum ($B=0.043T$) of the image in (b) to show the time response of the ballistic transport signal.  The lower trace was made in the same way for $B=-0.043T$.}
\end{figure}

  To further isolate the transport response, the raw data was filtered so that only those parts of the response which vary with magnetic field remain.  This removes the coupling signal that is largely independent of \textit{magnetic field} (a constant offset of each field scan).  Filtered scans are assembled in the greyscale plot of Fig.3b with time along the x-axis and magnetic field along the y-axis.  Here, the light areas correspond to electrons entering the collector.  The ballistic electron signal is quite pronounced in the form of horizontal white streaks at the focusing peak locations in magnetic field. This is further emphasized by the time averaged plot to the right of the main image. A weakly damped magnetoplasmon oscillation is also seen in the time response beginning at about $t=70ps$.  This resonance appeared in both positive and negative magnetic field and its frequency varied from $50GHz$ at zero field, to approximately $85GHz$ at $B=175mT$.  Although not displayed in the data of Fig.3b, these oscillations decayed away after approximately $250ps$.  A less pronounced set of similar but higher frequency oscillations can be seen from $t=20ps$ to $t=40ps$.  

The time response of the ballistic electrons travelling directly from the emitter to collector, displayed as the upper trace in Fig.3c, was obtained by averaging a $5mT$ wide horizontal slice at the first focusing maximum of the filtered response in Fig.3b at $B=0.043T$ (indicated by the arrow).  A similar slice of the filtered data, taken at $B=-0.043T$ is shown as the lower trace in Fig.3c.  We interpret the absence of any structure in the time response at this negative field to mean that the peaks in the ballistic electron response are due to changes in the emitter output, and not an artifact of the magnetoplasmon resonance modulating the signal.  

The first peak of the ballistic electron time response is centered at $t=52ps$, with an onset at about $30ps$.  Given the $50ps$ window before the arrival of waveguide reflections, the time-resolved ballistic electron signal for $t< 80ps$ ($50ps$ after the detection of the first electrons) will not be affected by the reflections.  The structure in the ballistic electron signal (Fig.3c) around $120ps$ and $170ps$ may arise from these waveguide reflections.  However, the first peak in Fig.3c occurs at too early a time to be affected and we attribute it to the ballistic electrons launched by the electrical pulse at $t=0$.

The width of the first ballistic electron peak in Fig.3c is approximately $35ps$, which is significantly broader than the initial electrical pulse.  Group velocity dispersion was calculated to be only $2ps$ for a $1ps$ pulse injected into the transport region.  A wide point contact produces an electron beam with approximately a $cos(\alpha)$ angular distribution  where $\alpha$ is the angle from perpendicular injection\cite{tribilliard}. Despite this, at fields meeting the focusing condition a high percentage of the electrons converge on the collector\cite{tribilliard}.  These electrons, launched at different angles, traverse paths of different lengths and result in a distribution of transit times.  A classical analysis\cite{tribilliard} was used to model the resulting time response to a $1ps$ emitter current pulse.  Assuming all carriers move at the Fermi velocity, and using the geometry of Fig.1b, a response width of approximately $18ps$ was predicted. While this simple model does not fully explain the width of the measured response, it nonetheless indicates that angular dispersion of the emitter point contact beam contributes heavily to the overall response width.

The modeled ballistic response was peaked at $28ps$ with an onset at $17ps$.  However, in Fig.3c the ballistic electron signal is instead seen to be peaked at $52ps$.  In Fig.2 and Fig.3b it is also clear that we see no magnetic focusing until $t>30ps$.  Precise agreement with the calculated $28ps$ base transit time is not expected since there are other possible delays.  For example, an additional delay of up to $10ps$ could be expected in order to traverse the point contacts (including a reduced velocity in the point contact).  At this time we do not have a full explanation of why these time-resolved measurements give a significantly longer delay than that expected from the simple Fermi velocity considerations.

\begin{figure}
\centering
\resizebox*{7.5cm}{!}{\includegraphics{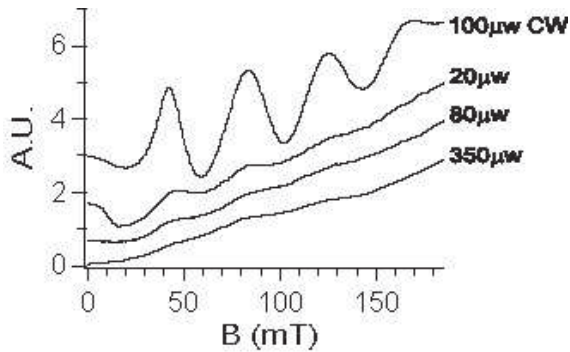}}
\caption{Comparison of reverse device operation using CW ($100\mu W$) and pulsed ($20,80,$ and $ 350\mu W$) optical excitation of the Auston switch.  Scans were offset and rescaled for clarity.}
\end{figure}

To investigate the effects of electrical pulse amplitude, the device of Fig.1 was also operated in the reverse direction by biasing the detector switch and exciting it with the pump beam to make it function as a pulse launcher.  In this configuration, the center waveguide conductor and one of the ground planes (connections 1 and 2 respectively in Fig.1c) form direct electrical connections to the focusing device and were used to measure the time averaged focusing signal.  Figure 4 compares the focusing spectrum measured using both continuous wave (CW) and pulsed laser excitation. A current of approximately $100nA$ was measured as being drawn from the DC voltage source at $100\mu W$ laser power (either CW or pulsed).  

In general, the focusing peaks in the pulsed case degraded both in contrast and amplitude at higher laser powers.  This is a signature of increased scattering due to electron heating\cite{EEHeating} and can be reproduced in similar magnetic focusing devices (using conventional CW current drive) by running them at high currents.  It was found that reducing the switch bias at fixed optical power had the same effect as lowering the optical power at fixed switch bias.  This confirmed that the heating effects were directly dependent on electrical pulse amplitude (a combination of switch bias and pump laser power) and are not arising from laser heating.

With the hot electron effects in mind, measurements were taken in the time-resolved configuration (Fig.1) at pump laser powers as low as $20\mu W$ (due to differences in switch efficiency, $300\mu W$ on the launch switch corresponded to approximately $80\mu W$ on the detection switch).  There was no change in the time response of the focusing signal as a function of launching power over this range.  This demonstrated that, compared to other broadening mechanisms, the emitter pulse voltage amplitude did not contribute heavily to the ballistic electron response of Fig.3c.

We have directly time-resolved ballistic electron transport in a transverse magnetic focusing device (fabricated from an $Al_{0.3}Ga_{0.7}As/GaAs$ heterostructure 2DEG) with $<5ps$ time resolution. Picosecond electrical pulses, delivered to the device by a waveguide, injected electrons through the emitter point contact and their arrival at the collector was time-resolved optoelectronically. A clear ballistic transport signal was observed along with magnetoplasmon oscillations in the device.  The measured transport time was $52ps$, nearly twice as long as the expected time-of-flight for ballistic electrons travelling at the Fermi velocity through the focusing orbit.  Additional delays inherent in the device have not been measured independently and the time difference cannot be completely explained at this time.  The measured response width of the ballistic electron signal was found to be approximately $35ps$.  After investigating various sources of response broadening, including group velocity dispersion and hot electron injection, the angular dispersion of the emitter point contact was found to contribute most heavily to the response width (accounting for approximately $18ps$ in calculations).  Other sources of response broadening have not yet been identified.

\begin{acknowledgments}
E.A. Shaner was supported by a National Defense Science and Engineering graduate fellowship.  This work was supported in part by the U.S. Army Research Office under grant \(\#\)DAAG55-98-1-0270.
\end{acknowledgments}

\bibliography{tmf}

\end{document}